\documentclass[aps,pre,print,onecolumn,amssymb,nobibnotes,longbibliography,unsortedaddress]{revtex4-2}
\usepackage{fancyhdr}
\usepackage{graphicx}
\usepackage{amsmath}
\usepackage{lipsum}
\usepackage{amsmath}
\usepackage[english]{babel}
\usepackage{graphicx}
\usepackage{dcolumn}
\usepackage{bm}
\usepackage[mathlines]{lineno}
\usepackage{setspace}
\usepackage{color}
\usepackage[utf8]{inputenc}
\usepackage[T1]{fontenc}
\usepackage{etoolbox}
\usepackage{hyperref}
\hypersetup{
    colorlinks=true,
    linkcolor=black,
    filecolor=black,
    urlcolor=black,
    citecolor=black,
    }
\usepackage{comment}
\usepackage{natbib}   
\usepackage{siunitx}
\usepackage[capitalise]{cleveref}
\usepackage{tikz}

\newcommand{\sym}[2]{\raisebox{0pt}[0pt][0pt]{\scalebox{#1}{#2}}}

\newcommand{\mystar}[2]{%
  \raisebox{2pt}[0pt][0pt]{%
    \scalebox{#1}{%
      \tikz[baseline=-0.6ex]{
        \fill[#2]
          (90:0.9em) -- (210:0.9em) -- (330:0.9em) -- cycle
          (270:0.9em) -- (30:0.9em) -- (150:0.9em) -- cycle;
      }%
    }%
  }%
}

\begin{document}

\title{Maximal spreading of impacting viscoelastic droplets}%

\author{Orr Avni}
\altaffiliation{These authors contributed equally to this work}
\affiliation{School of Engineering, Brown University, 184 Hope Street, Providence, RI 02912, USA}
\author{Dongyue Wang}
\altaffiliation{These authors contributed equally to this work}
\affiliation{School of Engineering, Brown University, 184 Hope Street, Providence, RI 02912, USA}
\author{Mithun Ravisankar}
\affiliation{Department of Mathematics, Mechanics division, University of Oslo, Oslo 0316, Norway}

\author{Roberto Zenit}
\email{roberto\_zenit@brown.edu}
\affiliation{School of Engineering, Brown University, 184 Hope Street, Providence, RI 02912, USA}

\date{\today} 
\begin{abstract}
{Droplet impact and spreading on solid substrates are well understood for Newtonian fluids, yet how viscoelasticity alone modifies the maximal spreading remains unclear.
 {Here, we isolate the role of viscoelasticity by performing impact experiments with Boger fluids of comparable viscosity and surface tension but different relaxation times.} 
For a wide range of conditions, viscoelastic droplets follow a similar behavior to Newtonian ones{, showcasing the same leading-order inertia--capillary scalings}; however, their maximal spreading diameter is significantly reduced compared with the Newtonian behavior when the Deborah number is of order unity.
 {This reduction is non-monotonic in the relaxation time: droplets with larger relaxation times recover Newtonian-like maximal spreading within experimental uncertainty. 
We rationalize these observations using an energy-balance model that includes a viscoelastic energy-storage term.}
The scaling argument obtained from this model explains the reported reduction in maximal spreading and identifies the range of fluid properties for which the strongest viscoelastic effects emerge.}

\end{abstract}

\maketitle
\thispagestyle{fancy}
\section{Introduction}\label{sec:intro}
Droplet impact on solid substrates is a canonical problem in fluid mechanics; it is a critical aspect in many engineering applications, ranging from inkjet printing and additive manufacturing to agricultural spraying and thermal control \citep{chandra_collision_1997,bonn_wetting_2009,lohse_fundamental_2022}.  
Despite its apparent simplicity, droplet impact couples inertia, viscosity, and capillary effects over a broad range of time and length scales; this interplay leads to complex dynamics and outcomes, including spreading, recoiling, bouncing, and splashing \citep{yarin_impact_1995,rioboo_time_2002,eggers_drop_2010}. 
Here, we focus on how viscoelasticity modifies the maximal spreading of impacting droplets.
For Newtonian liquids, extensive theoretical, numerical, and experimental work has established how the maximal spreading diameter $D_{\max}$ depends on the impact conditions, i.e., the droplet initial size $D_0$ and impact velocity $U_0$, as well as the fluid properties.  
These are commonly expressed in terms of the impact Weber, Reynolds, and Ohnesorge numbers, $\mathrm{We} = \rho U_0^2 D_0 / \sigma$, $\mathrm{Re} = \rho U_0 D_0 / \mu$, and $\mathrm{Oh}= \sqrt{\mathrm{We}}/\mathrm{Re}=\mu/\sqrt{\rho\sigma D_0}$, respectively, where $\rho$, $\mu$, and $\sigma$ are the liquid density, dynamic viscosity, and surface tension \citep{pasandidehfard_capillary_1996,eggers_drop_2010,ma_clinching_2024,sanjay_unifying_2025,liu_maximum_2025}.

Numerous scaling arguments predict two asymptotic regimes: viscous-limited and capillary-limited spreading.
In the viscous regime, where $\mathrm{We}\gg\mathrm{Re}^{1/2}$, the maximal diameter scales as either $\mathrm{Re}^{1/4}$  or $\mathrm{Re}^{1/5}$ \citep{pasandidehfard_capillary_1996,ma_clinching_2024}.
In the capillary regime, energy-conservation considerations lead to $\bar{D}_{\max}\sim \mathrm{We}^{1/2}$, whereas a momentum-balance argument yields $\bar{D}_{\max}\sim \mathrm{We}^{1/4}$.
\citet{laan_maximum_2014} proposed a semi-empirical relation that bridges these limiting cases by introducing the impact parameter $P=\mathrm{Re}^{-2/5}\mathrm{We}$ and a Padé-type correlation
\begin{equation}\label{eq:laan}
\bar{D}_{max}\mathrm{Re}^{-1/5} = \frac{\sqrt{P}}{1.24+\sqrt{P}},
\end{equation}
where $\bar{D}_{max}= D_{max}/D_0$.
More recently, \citet{sanjay_unifying_2025} reformulated the energy balance in terms of distinct impact and spreading stages, yielding a unified description of $D_{\max}$ across a wide range of impact conditions and clarifying the role of viscous dissipation on the spreading process.

In contrast to Newtonian fluids, the effect of non-Newtonian rheology on impact spreading is less understood. On the one hand, several studies have demonstrated that the spreading diameter of such droplets can deviate from Newtonian scalings \citep{an_maximum_2012,boyer_drop_2016}.  
Most recently, \citet{mobaseri_maximum_2025} showed that existing Newtonian correlations may be extended to incorporate shear-thinning and shear-thickening drops by using an effective shear rate to account for the average viscosity during spreading.  
On the other hand, less is known about how viscoelasticity modifies the maximal spreading of impacting droplets.  
Viscoelastic stresses are known to affect free-surface flows and capillary thinning \citep{bonn_viscoelastic_1997,dinic_macromolecular_2019,song_non-maxwellian_2023}, and to alter the oscillations of free droplets \citep{tamim_oscillations_2021}. 
In the related problem of capillary-driven spreading of sessile droplets, \citet{rostami_spreading_2024} showed that viscoelastic droplets may spread faster than Newtonian droplets with the same zero-shear viscosity and surface tension, and attributed this behavior primarily to shear thinning near the moving contact line. Conversely, \citet{agrawal_analysis_2019} showed, based on analytical considerations and experimental evidence, that polymeric viscoelasticity can reduce the capillary spreading velocity by increasing viscous dissipation within the droplet. These opposite trends highlight the complexity of decoupling shear-rate-dependent viscosity from elasticity.

Fluid elasticity in this context is usually quantified via the Deborah number, $\mathrm{De}=\lambda/\tau$, defined as the ratio of the fluid elastic relaxation time $\lambda$ to a characteristic flow timescale $\tau$.
Some studies report that, at large Deborah numbers $\mathrm{De}>10$, the spreading of non-Newtonian droplets is similar to that of Newtonian fluids \citep{laan_maximum_2014,gorin_universal_2022}, while other works find deviations from Newtonian behavior during retraction \citep{bergeron_controlling_2000,bartolo_dynamics_2007}.
Yet, studies that relate impact conditions and viscoelastic fluid properties to the maximal spreading over a wide range of parameters remain scarce.

 {Thus, this work aims to quantify how fluid elasticity modifies the maximal spreading of impacting Boger-fluid droplets and to identify the control parameters governing their departure from Newtonian behavior in the high-inertia regime, $\mathrm{Re}\gg1$ and $\mathrm{We}\gg1$.}  
\Cref{sec:exp} describes the experimental apparatus, which was designed to test the impact of Newtonian and viscoelastic Boger fluids with comparable viscosity and surface tension, but different relaxation times. \Cref{sec:spread} examines the spreading dynamics, including the characteristic spreading timescale, and establishes a Newtonian benchmark against which viscoelastic effects can be quantified.
Finally, \Cref{sec:viscoelasic_effects} develops an energy-balance model for the viscoelastic contribution to maximal spreading, compares its predictions with the experimental data, and relates the observed behavior to conventional viscoelastic dimensionless groups.

\section{Experimental setup and methods}\label{sec:exp}
To isolate the role of viscoelasticity on the spreading dynamics, a set of liquids with varying viscoelastic properties was prepared. Stock aqueous solutions of non-ionic polyacrylamide (PAAmm, Sigma-Aldrich) with molecular weight $M_w = 5 \times 10^6 \,\mathrm{g/mol}$ were first prepared and then diluted using different mixtures of glycerol (Nature's Oil, Vegetable Glycerin) and deionized water. The solutions were gently mixed with an overhead mixer for at least $12$ hours to ensure homogeneity. To minimize shear-dependent variations in viscosity, a small amount of magnesium sulfate salt ($0.025\,\mathrm{mol/L}$) was added to all solutions. To avoid polymer--polymer chain interactions and major variations in the zero-shear viscosity, the polymer concentration was kept below $300\,\mathrm{ppm}$ \citep{ravisankar_elastic_2025}.

\begin{figure}
    \centering
    \includegraphics[width=\linewidth]{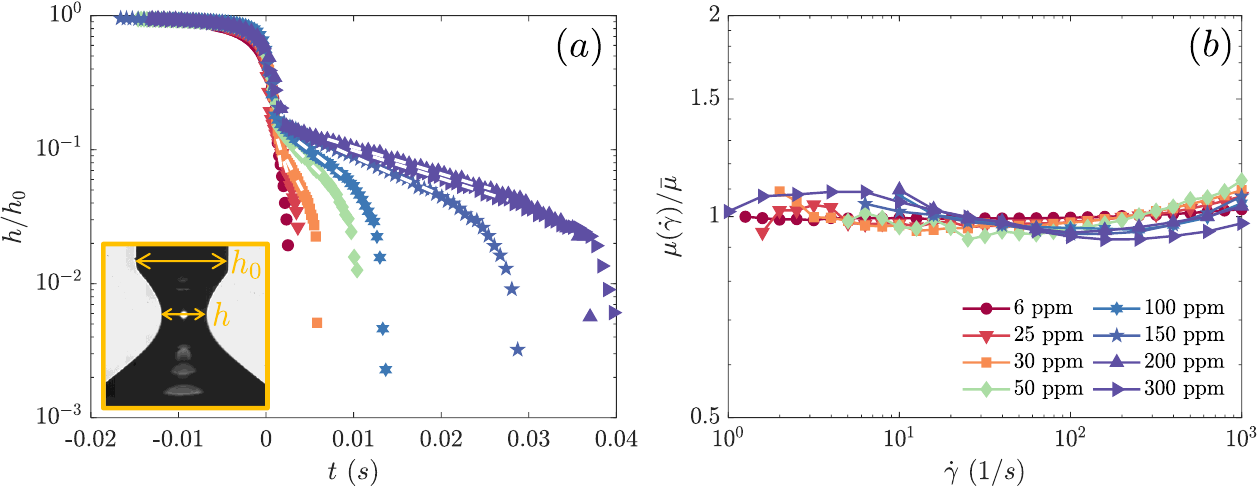}
    \caption{(a) Evolution of the minimum neck radius, $h$, normalized by the nozzle radius, $h_0$, during capillary thinning for fluids of different polymer concentrations. The time axis is shifted by the transition time $t_c$ between the inertio-capillary and elasto-capillary regimes. {The relaxation time, $\lambda$, is obtained from the slope of the dashed lines in the elasto-capillary regime, as predicted by \cref{eq:relaxation}.} Inset: definition of the neck radius $h$ and nozzle radius $h_0$. (b) Shear viscosity, normalized by its mean value $\bar{\mu}$, as a function of shear rate. The weak variation with shear rate indicates that shear-thinning is negligible for all tested solutions.}
    \label{fig:rheology}
\end{figure}

The shear viscosity, density, and surface tension of the fluids were measured using an ARES-G2 rheometer (TA Instruments), a density meter (Anton Paar), and a bubble-pressure tensiometer (KRUSS Scientific Instruments), respectively. The properties of all fluids used in the experiments are listed in \cref{tab:fluids}. The relaxation time, $\lambda$, was determined using an in-house dripping-onto-substrate method, following previous works \citep{dinic_extensional_2015,ravisankar_elastic_2025}. In this method, a pendant drop is brought into contact with a sessile drop, forming an unstable liquid bridge between the nozzle and the substrate. The subsequent capillary thinning of this bridge is recorded from the inertio-capillary stage ($t<0$) through the elasto-capillary regime ($t>0$) until breakup. By tracking the filament diameter $h(t)$ during the elasto-capillary stage, the relaxation time is obtained from \citet{dinic_extensional_2015}
\begin{equation}\label{eq:relaxation}
    \frac{h(t)}{h_0} = \left(\frac{G h_0}{2\sigma}\right)^{1/3}\exp\left(-\frac{t}{3\lambda}\right), \qquad t>0.
\end{equation}
where $h_0$ is the capillary diameter, and G is the polymer shear modulus. 
Representative thinning curves are shown in \cref{fig:rheology}(a); the corresponding values of $\lambda$ increase with polymer concentration.
The shear-thinning behavior is neglected since the reduction of viscosity is not significant over the relevant shear rates for our test fluids, as shown in \cref{fig:rheology}(b). In addition, the solvent viscosity was adjusted using water--glycerin mixtures so as to obtain fluids with comparable impact conditions while independently varying the relaxation time.
\begin{table}
\caption{Properties of the test fluids. Note that $\theta_e$ has two values, corresponding to paper and glass substrate, respectively.}
\label{tab:fluids}
\centering
\begin{ruledtabular}
\begin{tabular}{cc l c c c c c c}
Paper & Glass & Polymer & Glycerin &
$\mu$ (mPa\,s) & $\rho$ (kg/m$^{3}$) & $\sigma$ (mN/m) & $\lambda$ (ms) & $\theta_e$ (degrees)\\
\hline
\multicolumn{8}{l}{\textit{Viscoelastic fluid set}} \\
\sym{1.5}{$\bullet$}           & \sym{1.5}{$\circ$}        & 6 ppm   & 80\% & 38.5 & 1193.4 & 77.3 & 4.6 & 128.7, 72.5 \\
\sym{1.1}{$\blacktriangledown$}& \sym{1.1}{$\triangledown$}& 25 ppm  & 50\% & 5.13 & 1121.4 & 69.7 & 5.7 & 114.2, 70.5\\
\sym{0.9}{$\blacksquare$}      & \sym{0.9}{$\square$}      & 30 ppm  & 50\% & 5.51 & 1131.2 & 71.6 & 8.0 & 122.2, 69\\
\sym{1.0}{$\blacklozenge$}     & \sym{1.0}{$\lozenge$}     & 50 ppm  & 0\%  & 1.39 & 999.5  & 71.7 & 17.2 & 132.8, 70.2\\
\mystar{0.45}{black}           &                            & 100 ppm & 0\%  & 1.68 & 997.2  & 73.8 & 61.9 & 132.7\\
\sym{0.9}{$\bigstar$}          &                            & 150 ppm & 0\%  & 1.60 & 997.6  & 73.1 & 70.3 & 125.7\\
\sym{1.1}{$\blacktriangle$}    &                            & 200 ppm & 0\%  & 1.84 & 992.6  & 73.2 & 83.6 & 132.2\\
\sym{1.0}{$\blacktriangleright$}&                           & 300 ppm & 0\%  & 1.92 & 996.8  & 73.0 & 79.7 & 133\\
\hline
\multicolumn{8}{l}{\textit{Newtonian set}} \\
\sym{1.0}{$\times$}            & \sym{1.0}{$+$}            & 0 ppm   & 80\% & 40.5 & 1199.6 & 77.3 & 0 & 124, 83.1\\
\sym{1.0}{$\times$}            &                            & 0 ppm   & 30\% & 2.37 & 1064.2 & 71.6 & 0 & 117.1\\
                               & \sym{1.0}{$+$}            & 0 ppm   & 0\%  & 1.05 & 998.4  & 71.2 & 0 & 66.6\\
\end{tabular}
\end{ruledtabular}
\end{table}

The solutions were used to form droplets using a syringe pump, pushing the liquid through a straight stainless-steel needle. By using different needle tips, droplets of varying initial diameters were obtained, from $D_0 = \qtyrange{2.5}{4.0}{\mm}$. 
{The needle was positioned at different heights above a flat, leveled stage, leading to impact velocities in the range $U_0=\qtyrange{1.0}{2.5}{\m/s}$, corresponding to impact Reynolds numbers in the range $\mathrm{Re}=O\left(10^2 - 10^4\right)$ and impact Weber numbers in the range $\mathrm{We}=O\left(10^1 - 10^2\right)$.}
Two types of substrates were placed on top of the stage: microscope-grade glass (AmScope, BS-72P) and hydrophobic paper sheets (Strathmore, Watercolor, 400 Series), and were replaced after each experiment. 
The impact of the droplet was recorded by three high-speed cameras simultaneously. 
A wide-angle view camera (Photron SA5, $5000\,\mathrm{fps}$) was used to measure the droplet diameter and impact speed at the moment of impact, as illustrated in the left-hand inset of \cref{fig:exp}. 
The spreading process was captured by a narrow-angle view camera (Photron NOVA R-5, $40000\,\mathrm{fps}$); selected subsequent snapshots are presented in \cref{fig:exp}. 
With the camera's high recording rate and resolution ($2048\times64$ pixels), we were able to determine the maximal spreading diameter and spreading time at high temporal and spatial resolution. 
Seeking to minimize the spread of the results and reduce the uncertainty, impact events in which the rim of the droplet exhibited strong irregularity, i.e., splashing and the onset of fingers, were discarded and repeated. 
The irregularity was identified using an isometric view camera (Photron SA3, $5000\,\mathrm{fps}$) and was limited to within $5\%$ of the equivalent diameter. The equilibrium contact angle, $\theta_e$ was measured from images of the sessile droplets remaining after the collision, using image processing tools.

\section{Spreading dynamics}\label{sec:spread}
\begin{figure}
  \centering
  \includegraphics[width=\linewidth]{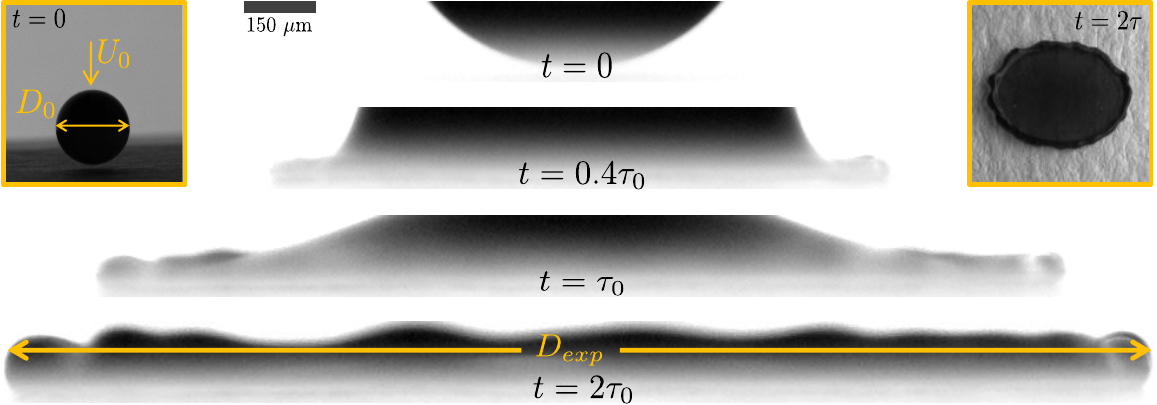}
  \caption{Successive snapshots of a single droplet impacting on a hydrophobic paper substrate, progressing from impact (top) to maximum spread (bottom). 
  Time is given in terms of the impact timescale, $\tau_0 = D_0/U_0$. 
  Top left: representative frame from the wide-angle camera, used to determine the impact conditions. 
  Top right: representative frame from the isometric-view camera, recording the droplet topology.}
\label{fig:exp}
\end{figure}
A representative impact event recorded by the camera array is shown in \cref{fig:exp}.
As it impacts the substrate, the droplet flattens and radially expands over the substrate, while its shape evolves into a `pizza-like' shape with a prominent rim \citep{wildeman_spreading_2016}. 
This behavior is consistent with the impact conditions ($\mathrm{We} > 30$, $\mathrm{Oh} < 0.1$), for which energy dissipation is predominantly confined to a thin layer near the substrate \citep{sanjay_unifying_2025}. 
For all fluids and substrates used here, the droplets remain attached, and no rebound is observed, although the diameter recedes and oscillates after reaching its maximum; we therefore focus here on the maximal spreading diameter, $D_{\mathrm{max}}$, as it is a strong indicator of the eventual outcome of the impact event. 

\begin{figure}
    \centering
    \includegraphics[width=\linewidth]{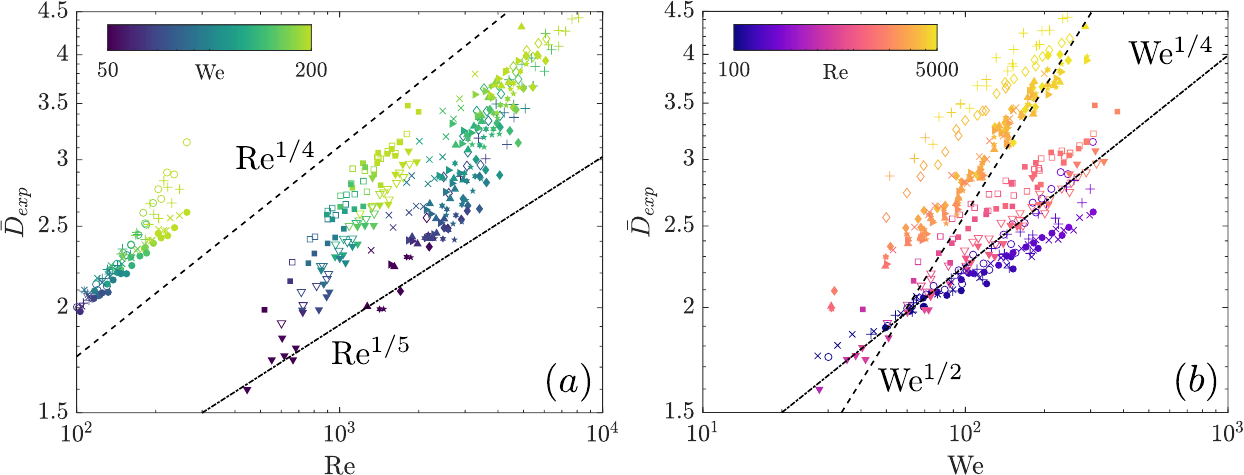}
\caption{Measured maximal spreading diameter $\bar{D}_{\mathrm{exp}}$ of single droplets of various Newtonian and viscoelastic fluids, detailed in \cref{tab:fluids}.
(a) $\bar{D}_{\mathrm{exp}}$ versus the impact Reynolds number $\mathrm{Re}$; dashed and dotted lines indicate $\mathrm{Re}^{1/4}$ and $\mathrm{Re}^{1/5}$ scalings, respectively. 
Marker color denotes the impact Weber number $\mathrm{We}$. 
(b) $\bar{D}_{\mathrm{exp}}$ versus the impact Weber number; dashed and dotted lines indicate $\mathrm{We}^{1/4}$ and $\mathrm{We}^{1/2}$ scalings, respectively. 
Marker color denotes the impact Reynolds number.} 
    \label{fig:ReWe}
\end{figure}

To evaluate the differences between viscoelastic and Newtonian impacts, we first examine the normalized measured spreading diameter $\bar{D}_{\mathrm{exp}}=D_{\mathrm{max}}/D_0$ as a function of the impact Reynolds and Weber numbers in \cref{fig:ReWe}. 
Within the experimental scatter, no systematic difference between glass and paper substrates is observed, indicating that substrate effects are weak in the present parameter range. 
{Qualitatively, both Newtonian and viscoelastic fluids exhibit the same maximal-spreading scaling behavior, in general agreement with the models discussed in \cref{sec:intro}.}
For large $\mathrm{We}$, the scaling with $\mathrm{Re}$ tends towards $\bar{D}_{\max}\propto\mathrm{Re}^{1/4}$, as evident in \cref{fig:ReWe}(a). 
At the other end of the scaling range, for large $\mathrm{Re}$, a $\mathrm{We}^{1/2}$ scaling emerges in \cref{fig:ReWe}(b), again consistent with the limits of our model, which supports the view that an energy-balance framework is appropriate for estimating the maximal spreading under our experimental conditions. 
Overall, the scalings in \cref{fig:ReWe} show that viscoelastic droplets obey the same leading-order inertia–capillary asymptotic behavior as Newtonian droplets; still, they do not yet isolate the viscoelastic effects.  

\begin{figure}
    \centering
    \includegraphics[width=\linewidth]{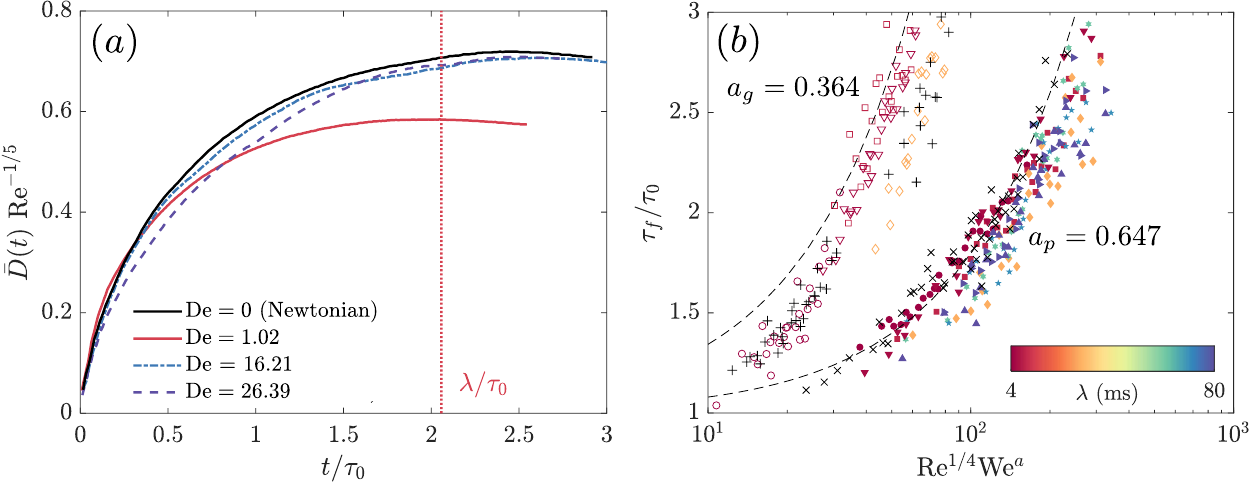}
\caption{(a) Temporal evolution of the normalized spreading diameter, $\bar{D}_{\exp}\mathrm{Re}^{-1/5}$, for four representative droplets (0 ppm, 25 ppm, 100 ppm, and 200 ppm of PAAmm) with Deborah numbers $0$, $1.02$, $16.21$, and $26.39$, respectively, all at an impact parameter range $P = 6.75 \pm 0.05$. 
The vertical dashed line marks the dimensionless relaxation time $\lambda/\tau_0$ for the case $\mathrm{De}=1.02$; for the other viscoelastic cases, $\lambda/\tau_0$ lies outside the spreading timescale. 
(b) The ratio between measured spreading time $\tau_f$ and the inertial timescale $\tau_0$ as a function of $\mathrm{Re}^{1/4}\mathrm{We}^a$, where $a$ is a power-law coefficient fitted for each type of substrate. The dashed lines denote the proposed correlation, \cref{eq:corr}}.
    \label{fig:dyn}
\end{figure}

Nonetheless, when considering the rescaled maximal spreading diameter $\bar{D}\,\mathrm{Re}^{-1/5}$ following \citet{laan_maximum_2014}, the variations in the spreading dynamics begin to emerge.
{One would expect that, in the absence of viscoelasticity, all spreading events with similar impact parameter $P=\mathrm{We}\mathrm{Re}^{-2/5}$ would unfold in the same manner.}
We consider the time evolution of the spread in four cases of impact; in all of them, the impact parameter lies within a narrow range of $P = 6.75 \pm 0.05$, but only varying the fluid elastic relaxation time. 
To keep the scaling equivalent, the time is normalized with respect to the inertial timescale, $\tau_0=D_0/U_0$.
\Cref{fig:dyn}(a) reveals that each impact begins with a short contact phase ($t/\tau_0 \lesssim 0.1$) during which the droplet first hits the substrate and flattens. 
No significant variation between the impacts is evident at this stage.
Overall, the droplet spreads approximately as $D \propto \sqrt{t}$, consistent with the canonical inertial–capillary regime reported in previous experimental and numerical studies \citep{pasandidehfard_capillary_1996,eggers_drop_2010}.
Following the initial contact period, a significant difference between the cases is observed for $t/\tau_0 \ge 0.5$: when the Deborah number is around unity, spreading is significantly hindered compared with the Newtonian case.  
The maximal (rescaled) diameter is noticeably lower, and the spreading time, defined as the time between initial contact and maximal spreading, is shorter; other viscoelastic droplets do not exhibit a substantial deviation from the Newtonian dynamics. Although they have higher polymer concentration and thus a higher Deborah number, both their diameter evolution and spreading time remain similar to those in the Newtonian case; the polymer addition affects the dynamics only via the change in the viscosity. 

It is important to note that we define here the Deborah number as the ratio between the elastic relaxation time and the measured spreading time $\mathrm{De}=\lambda/\tau_f$. 
The ratio between the spreading time and the inertial timescale, while still of the order of unity, varies in the range of 1 to 3, as evident in \cref{fig:dyn}(b). 
Interestingly, the upper limit roughly corresponds to the spreading time obtained by \citeauthor{pasandidehfard_capillary_1996}~\citep{pasandidehfard_capillary_1996}, $\tau_f/\tau_0=8/3$, obtained assuming potential flow within the lamella.
This result also reflects our realization that a single timescale value might not accurately reflect the spreading time, making the Deborah number an intrinsic parameter that may not be predicted a priori.
Nevertheless, the ratio between the spreading and inertial timescales was found to collapse as a function of $\mathrm{Re}^{1/4}$ and a power law of $\mathrm{We}$ which depends on the substrate type. As such, we correlated this ratio using the function
\begin{equation}\label{eq:corr}
    \frac{\tau_f}{\tau_0} = 1+b\left(\mathrm{Re}^{1/4}\mathrm{We}^a\right);
\end{equation}
for a hydrophobic paper substrate, it was found that $a_p=0.647$ and $b_p=0.0079$ yield the best agreement, while for a glass substrate, it is  $a_g=0.364$ and $b_g=0.0034$; the corresponding functions, alongside the measured time ratios, are presented in \cref{fig:dyn}(b).

\begin{figure}
    \centering
    \includegraphics[width=0.5\linewidth]{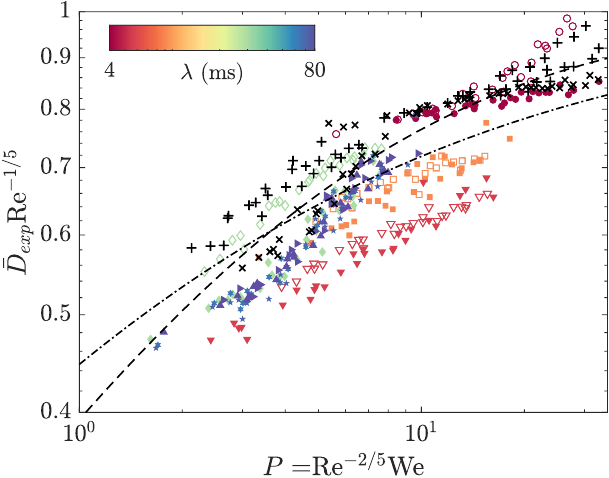}
\caption{{Maximal normalized spreading, $\bar{D}_{\exp}\mathrm{Re}^{-1/5}$, versus the impact parameter $P=\mathrm{We}\mathrm{Re}^{{-2/5}}$ for single impacting droplets of various Newtonian and viscoelastic fluids, detailed in \cref{tab:fluids}. Marker color indicates the fluid elastic relaxation time $\lambda$; Newtonian fluids with $\lambda\rightarrow0$ are shown in black. The Padé correlation of \citet{laan_maximum_2014} (dash-dotted) and a second-order fit to the present Newtonian data \cref{eq:pade_new} (dashed) are shown for comparison. }}
    \label{fig:Bonn}
\end{figure}

We may now test whether the single case where viscoelastic effects are maximized at $\mathrm{De}\approx1$ holds across the full set of impact events. 
\Cref{fig:Bonn} shows the rescaled maximal diameter as a function of the impact parameter for all of the tested fluids. 
The Newtonian data (red symbols) collapse onto a single, well-defined branch, whereas several viscoelastic fluids deviate from this trend.  
In particular, two fluids with relaxation times in the range $\lambda=6$--$10\,\mathrm{ms}$ exhibit the most significant deviations from the spreading of Newtonian droplets across all impact conditions. However, as we increase the elastic relaxation time, the spreading behavior gradually reverts towards the Newtonian case; the deviation from Newtonian behavior is therefore a non-monotonic function of the elastic relaxation time $\lambda$. 

To quantify the extent of this deviation and its dependence on the fluid properties and impact conditions, we first require a reliable benchmark correlation for the maximal spreading of Newtonian droplets under our experimental conditions.  
We therefore compare in \cref{fig:Bonn} the Newtonian data to the Padé approximant proposed by \citet{laan_maximum_2014}, given in \cref{eq:laan}; however, it is found to underestimate the rescaled maximal spreading in our parameter range.  
To determine whether this discrepancy is specific to the Laan correlation, we further compared our Newtonian data with several previously published correlations; the complete comparison is provided in the Appendix.
These comparisons motivate the use of an alternative benchmark correlation better suited to our experimental parameter range.
We therefore adopt a second-order Padé approximant, similar in form to that proposed by \citet{mobaseri_maximum_2025},  
\begin{equation}
    \bar{D}_{\mathrm{cor}}\mathrm{Re}^{-1/5} = \frac{0.37P + \sqrt{P}}{0.37P + \sqrt{P} + 2.1}.
    \label{eq:pade_new}
\end{equation}

\begin{figure}
    \centering
    \includegraphics[width=0.5\linewidth]{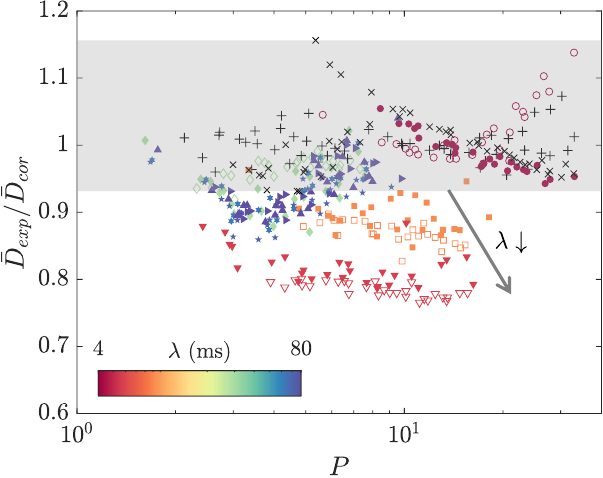}
    \caption{Deviation of the measured spreading diameter from the expected value at given impact conditions, $\bar{D}_{\exp}/\bar{D}_{\mathrm{cor}}$, as a function of the impact parameter $P=\mathrm{Re}^{{-2/5}}\mathrm{We}$. The shaded area visualizes the spreading of the Newtonian results.} 
    \label{fig:corr_p}
\end{figure}

Using \cref{eq:pade_new}, we can now directly compare the measured maximal spreading to the Newtonian prediction at the same impact conditions by calculating the ratio $\bar{D}_{\exp}/\bar{D}_{\mathrm{cor}}$.  
As evident in \cref{fig:corr_p}, the correlation captures the Newtonian data reasonably well, estimating the maximal spreading diameter within approximately $\pm 10\%$ across the tested impact conditions.  
In this representation, the effects of viscoelasticity emerge clearly.
While the correlation prediction for most of the Newtonian impacts lies within the 10\% accuracy range, the introduction of polymers leads to a reduction of up to 25\%, for a relaxation time of $\lambda\approx 6$ ms, which corresponds to De$=1$--2. 
The deviation is clearly non-monotonic with $\lambda$, but is consistent across a wide range of impact parameters. 

At the edge of our parametric sweep, for relatively high polymer concentration, we again observe some decrease in the spreading diameter, especially for lower impact parameters.
{Interestingly, the deviation becomes insignificant again for higher impact parameter values. This trend may be explained by polymer migration away from the contact line due to the high shear rates experienced locally near the contact line. Even though the tested fluids did not exhibit any shear-thinning behavior for shear rates up to $O(10^3)$, during the impact event, especially for high-momentum impacts, the local shear rates may be of the order of $O(10^5)$. Recently, \citet{mobaseri_inertia-dilatancy_2026} illustrated similar behavior for shear-thickening fluids, exhibiting unexpected rheological behavior due to high shear rates during drop impact. However, the decrease observed for low impact parameters does not lie markedly outside the correlation uncertainty range and, in any case, is lower than that observed for slightly viscoelastic fluids, regardless of any possible polymer migration.}
In the following section, we rationalize this deviation and define the region in which the current models fail to describe the maximal spreading of viscoelastic droplets.

\section{The effect of viscoelasticity on the maximum spreading diameter}\label{sec:viscoelasic_effects}

 {The experimental observations presented in the previous section showcased how viscoelasticity can affect the spreading of an impacting droplet, although the conditions under which this effect becomes significant are not immediately evident.
To rationalize these observations and predict the regime in which viscoelastic effects become important, we consider an energy balance  that accounts for viscous dissipation, surface-energy changes, and elastic energy storage.}
\subsection{Conservation of Energy}
Following classical energy-balance arguments \citep{clanet_maximal_2004,sanjay_unifying_2025}, and assuming that no initial kinetic energy of the droplet remains at the moment of maximum spreading, we have
\begin{equation}\label{eq:S_balance}
E_0 = \Delta E_\sigma + \Delta E_\mu + \Delta E_{\lambda},
\end{equation}
where $E_0$ is the initial kinetic energy, $\Delta E_\sigma$ is the change in surface energy, $\Delta E_\mu$ is the energy dissipated by viscosity, and $\Delta E_{\lambda}$ is the elastic energy stored in the fluid. 
The initial kinetic energy is estimated as
\begin{equation} \label{eq:balance}
E_0 = \frac{1}{2}\rho V_D U_0^2,
\end{equation}
where $V_D=\pi D_0^3/6$ is the initial droplet volume and $U_0$ is the impact velocity. 

The change in surface energy is expressed as
\begin{equation}
\Delta E_\sigma =
\frac{\pi}{4}D_{\max}^2 \left(\sigma_{SL}-\sigma_{SA}\right)
+\frac{\pi}{4}\sigma_{LA}D_0^2\left(\bar{D}^2-4\right),
\end{equation}
where $\bar{D}=D_{\max}/D_0$; the subscripts $S, L$ and $A$ in the surface tension, $\sigma$, refer to interfacial tension values for solid, liquid and air combinations, respectively.
 {Using the Young--Dupr\'e relation, $\sigma_{SA}-\sigma_{SL}=\sigma_{LA}\cos\theta_d$ \cite{degennes1985}}, this expression becomes
\begin{equation} \label{eq:E_sigma}
\Delta E_\sigma =
\frac{\pi}{4}\sigma D_0^2
\left[\bar{D}^2\left(1-\cos\theta_d\right)-4\right],
\end{equation}
 {where $\theta_d$ is the dynamic contact angle and $\sigma\equiv\sigma_{LA}$ is the liquid--air surface tension.}

The viscous dissipation is estimated as
\begin{equation}
\Delta E_{\mu} = \phi V_{\mu}\tau_f,
\end{equation}
where $V_{\mu}$ is the sheared liquid volume, $\tau_f$ is the spreading time and $\phi$ is the volumetric viscous dissipation rate, approximated as  {\citep{pasandidehfard_capillary_1996}}
\begin{equation}
\phi \approx \mu\left(\frac{U_0}{\delta}\right)^2.
\end{equation}
where $\delta$ is the viscous boundary-layer thickness.
The elastic energy stored within the sheared liquid volume is estimated as
\begin{equation}
\Delta E_{\lambda} \approx V_{\mu}G\gamma^2,
\end{equation}
where $V_{\mu}$ is the sheared liquid volume, $G$ is the time-dependent elastic modulus, and $\gamma$ is the characteristic shear strain. Assuming a simple Maxwellian response \cite{macosko1994}, the effective modulus over the spreading time can be estimated as {
\begin{equation}
G(\tau_f) \approx
G_0\exp\left(-\frac{\tau_f}{\lambda}\right)
=
\frac{\mu}{\lambda}\exp\left(-\frac{\tau_f}{\lambda}\right)
=
\frac{\mu}{\tau_f\mathrm{De}}
\exp\left(-\frac{1}{\mathrm{De}}\right),
\end{equation}}
where $\lambda=\mu/G_0$ is the relaxation time, $\tau_f$ is the measured spreading time, and $\mathrm{De}=\lambda/\tau_f$. Assuming that the characteristic shear rate remains approximately constant throughout the spreading process, the shear strain is estimated as
\begin{equation}
\gamma \approx \tau_f \dot{\gamma}
\approx \tau_f\frac{U_0}{\delta},
\end{equation}
where $\delta$ is the viscous boundary-layer thickness. Therefore,
\begin{equation}
\Delta E_{\lambda}
\approx
\frac{\mu V_{\mu}\tau_f U_0^2}{\mathrm{De}\,\delta^2}
\exp\left(-\frac{1}{\mathrm{De}}\right).
\end{equation}
The viscous and elastic contributions can be readily combined to write
\begin{equation}
\Delta E_{\mu+\lambda}
=
\mu V_{\mu}\tau_f
\left(\frac{U_0}{\delta}\right)^2
\left[
1+\frac{1}{\mathrm{De}}\exp\left(-\frac{1}{\mathrm{De}}\right)
\right].
\end{equation}
This expression can be simplified further by assuming that the spreading time can be approximated by $\tau_f\propto D_0/U_0$. Assuming that only a fraction of the droplet volume is sheared during the impact, the sheared liquid volume can be estimated as
\begin{equation}
V_\mu=\frac{\pi}{4}D_0^2\bar{D}^2\delta.    
\end{equation}
In turn, $\delta$ can be estimated following a classical boundary-layer estimate \cite{schlichting2016} to be $\delta\approx D_0\mathrm{Re}^{-1/2}$. Note that this estimation of $\delta$ is based on the classical similarity solution for boundary layer flows, valid only for large Re (inertia-dominated flows).
Hence, we can write
\begin{equation} \label{eq:E_mu_lambda}
\Delta E_{\mu+\lambda}
=
\frac{\pi}{4}\sigma D_0^2\bar{D}^2
\left[
\frac{4\mathrm{We}}{3\sqrt{\mathrm{Re}}}
\right]
\left(1+\alpha\right).
\end{equation}
where 
\begin{equation}
\alpha(\mathrm{De}) =
\frac{1}{\mathrm{De}}\exp\left(-\frac{1}{\mathrm{De}}\right)
\end{equation}
encapsulates the effect of viscoelasticity. 

Finally, substituting expressions \cref{eq:E_sigma} and \cref{eq:E_mu_lambda} into \cref{eq:balance}, we obtain an expression for the maximum spreading diameter:
\begin{equation} \label{eq:S_Dmax}
\bar{D}
=
\sqrt{
\frac{\mathrm{We}+12}
{3(1-\cos\theta_d)+4(1+\alpha)\mathrm{We}\mathrm{Re}^{-1/2}}
}.
\end{equation}
This expression suggests a general correlation of the form
$\bar{D}\propto\mathrm{We}^a\mathrm{Re}^{b}$, as discussed in the Section III. Notably, for $\alpha=0$, \cref{eq:S_Dmax} reduces to the Newtonian case. 

To further isolate the effect of viscoelasticity, we consider the maximal spreading diameter ratio between the viscoelastic case and a Newtonian case under exactly the same impact conditions. Thus, one may rewrite
\begin{equation}\label{eq:D_max}
\frac{\bar{D}_{VE}^2}{\bar{D}_{N}^2}
=
\left(1+\frac{\alpha}{1+{1}/{K}}\right)^{-1},
\end{equation}
where
\begin{equation}
K =
\frac{4\mathrm{We}}
{3\sqrt{\mathrm{Re}}(1-\cos\theta_d)}.
\end{equation}
is a single parameter accounting for both the impact conditions and the substrate wettability. 
The simplicity of this expression allows us to identify the relevant conditions affecting the maximum spreading diameter.
It should be noted that \cref{eq:D_max} indicates that the value of the dynamic contact angle dictates the viscoelastic response only when We/Re$^{1/2}$ is of order of unity or smaller, resulting in small $K$.  {Although the dynamic contact angle, $\theta_d$, cannot be measured directly in our experiment, its value can inferred from the equilibrium contact angle $\theta_e$ using Hoffman's relation \citep{hoffman1975}:
\begin{equation}
\theta_d^3\approx\theta_e^3+72\frac{\mathrm{We}}{\mathrm{Re}}.
\end{equation}
Hence, for the range of Re and We considered here, $K$ ranges from 10 to 300 approximately, making the dependence on $\theta_d$ weak. The fact that similar spreading behavior is observed for paper and glass substrates, further confirms that wettability effects, while generally significant, do not play a relevant role in the considered experimental conditions.}

\subsection{Comparison with experimental data} 

\begin{figure}
\centering
\includegraphics[width=0.5\linewidth]{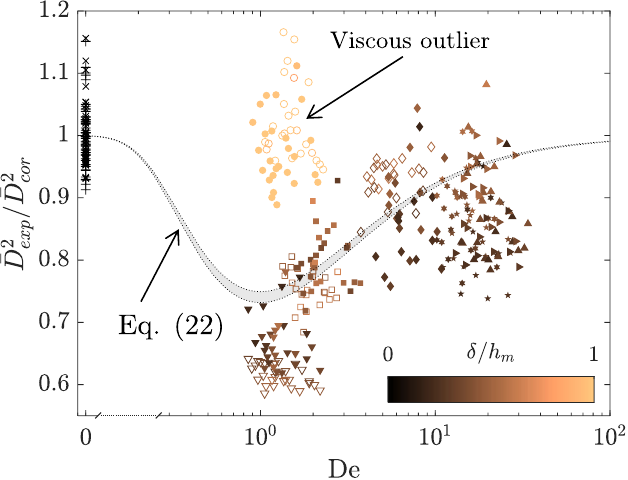}
\caption{Deviation of the squared measured spreading diameter from the expected value at given impact conditions, $\bar{D}^2_{\exp}/\bar{D}^2_{\mathrm{cor}}$, as a function of $\mathrm{De}$.
For reference, all the Newtonian data are plotted at $\mathrm{De}=0$.
Marker color indicates the ratio of viscous boundary-layer thickness to maximal film thickness, $\delta/h_m$.
The gray shaded region represents the range predicted by the energy-balance model for the reduction in maximal spreading due to viscoelastic effects (\cref{eq:D_max}) over the experimental range $10<K<300$.}
\label{fig:alpha}
\end{figure}

To test the prediction of \cref{eq:D_max}, we show in \cref{fig:alpha} the ratio $\bar{D}^2_{\exp}/\bar{D}^2_{\mathrm{cor}}$ as a function of the Deborah number, $\mathrm{De}$.
Consistent with the single case presented in \cref{fig:dyn}(a), and despite the scatter in the data, it is evident that the deviation from the Newtonian correlation is maximal when $\mathrm{De}=O(1)$. 
For $\mathrm{De}>10$, the effect of viscoelasticity becomes small, and the data approach the Newtonian value. 
This trend is  {agrees} with the scaling prediction from \cref{eq:D_max}, providing a physical basis as for why the spreading is affected most strongly when $\mathrm{De}\approx1$. 
The gray shaded region in \cref{fig:alpha}, corresponding to the experimentally relevant range $10<K<300$, captures the overall trend of the present data. 
The relatively narrow range of predictions also supports the observation that substrate type has only a weak effect on the maximal spreading under the present conditions.  {In general, the trend observed in our data and analysis is consistent with previous studies \citep{gorin_universal_2022}, which also reported Newtonian-like maximal spreading for $\mathrm{De}=10$--$40$.}

One set of measurements, however, lies systematically above the main viscoelastic branch at $\mathrm{De}\sim1$. Although the impact conditions lie within the range where a deviation is otherwise observed, this fluid, which is the most viscous fluid in our set (first row of \cref{tab:fluids}), exhibits spreading that is nearly identical to that of the Newtonian fluids.
This corresponds to the only sample with a relatively high Ohnesorge number, $\mathrm{Oh}\approx0.1$.
Under such conditions, the viscous boundary-layer thickness $\delta\approx\sqrt{\nu\tau_f}$ is estimated to exceed the measured droplet thickness at the point of maximal spread, $h_m$, such that $\delta/h_m\gtrsim1$.
The data in \cref{fig:alpha} are color-coded according to $\delta/h_m$, from dark to light for small to large values. When the boundary-layer thickness is no longer small compared with $h_m$, viscous dissipation becomes increasingly important and may mask the elastic effect encapsulated by $\alpha(\mathrm{De})$. 
Thus, even if elastic energy is stored, it is subsequently dissipated, and the fluid exhibits dynamics closer to those of a Newtonian fluid.
The present energy-balance model is therefore expected to apply when $\delta/h_m<1$, corresponding approximately to $\mathrm{Oh}\ll1$.

\subsection{Relation to viscoelastic dimensionless groups}
\begin{figure}
    \centering
    \includegraphics[width=\linewidth]{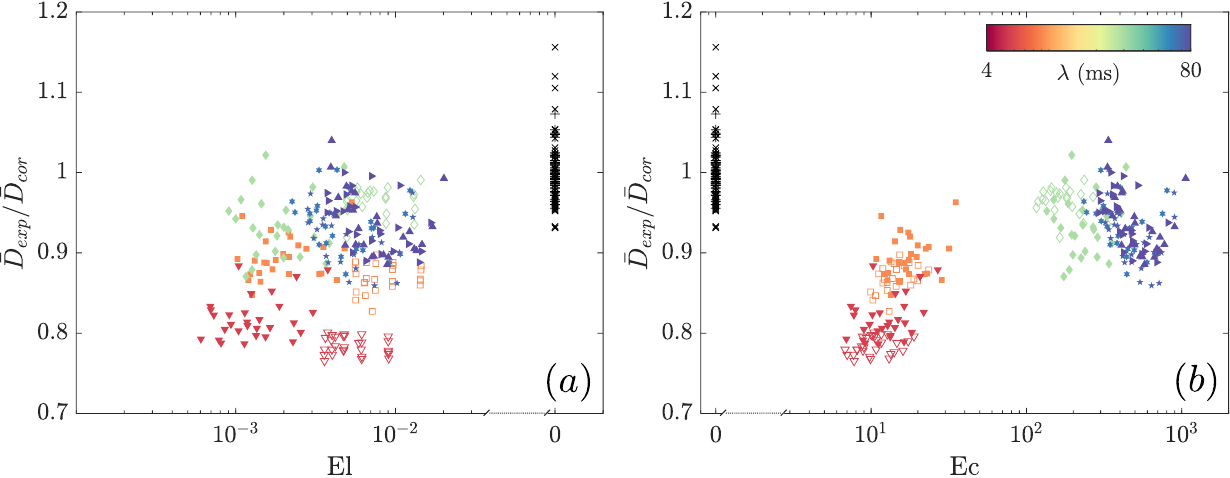}
    \caption{Deviation of the measured maximal spreading from the Newtonian correlation, $\bar{D}_{\exp}/\bar{D}_{\mathrm{cor}}$, plotted against viscoelastic control parameters. 
    Marker color indicates the fluid relaxation time $\lambda$. 
    (a) $\bar{D}_{\exp}/\bar{D}_{\mathrm{cor}}$ as a function of the elasticity number $\mathrm{El} = \mathrm{De}/\mathrm{Re}$. 
    (b) $\bar{D}_{\exp}/\bar{D}_{\mathrm{cor}}$ as a function of the elasto-capillary number $\mathrm{Ec} = \mathrm{De}\mathrm{Re}/\mathrm{We}$.}
    \label{fig:VE_ND}
\end{figure}

To further delineate the region in which viscoelasticity affects the maximal spreading, we examine the deviation from the Newtonian correlation, $\bar{D}_{\exp}/\bar{D}_{\mathrm{cor}}$, in terms of well-established dimensionless groups: the elasticity and elasto–capillary numbers. 
\Cref{fig:VE_ND}(a) shows $\bar{D}_{\exp}/\bar{D}_{\mathrm{cor}}$ plotted against the elasticity number $\mathrm{El}=\mathrm{De}/\mathrm{Re}$; over the examined range, $\mathrm{El}\sim10^{-3}$--$10^{-2}$, the data exhibit substantial scatter and no clear trend, indicating that $\mathrm{El}$ alone does not parameterize the viscoelastic correction under the present impact conditions.  
In contrast, when the same data are plotted as a function of the elasto–capillary number $\mathrm{Ec}=\mathrm{De}\mathrm{Re}/\mathrm{We}$ in \cref{fig:VE_ND}(b), a clear separation of scales emerges.  
For high $\mathrm{Ec}$ values ($10^2$--$10^3$), no significant viscoelastic effects are observed; however, as $\mathrm{Ec}$ decreases towards order $10^2$, the ratio $\bar{D}_{\exp}/\bar{D}_{\mathrm{cor}}$ decreases, indicating a growing reduction of the maximal spread relative to the Newtonian prediction. 
These trends suggest that measurable deviations in maximal spreading arise when elastic stresses become comparable to capillary stresses in the spreading lamella. 
Although not attainable in the current experimental setup, one might postulate, based on the observed trend, that elastic effects peak at $\mathrm{Ec}\approx 1$.
This agrees with the general trend shown by \citet{rostami_spreading_2024} for viscoelastic droplet wetting, although in their case the effect was attributed primarily to shear thinning near the moving contact line, whereas the present work concerns impact-driven maximal spreading of Boger-like fluids.

\section{Conclusions}\label{sec:conclusions}

The effect of viscoelasticity on droplet spreading upon impact remains difficult to isolate. 
To clarify the underlying physical mechanisms governing the spreading dynamics, we conducted impact experiments to quantify how fluid elasticity modifies the maximal spreading of impacting droplets and to identify the parameters governing departures from Newtonian behavior.
Experiments were performed using dilute polyacrylamide (PAAmm) solutions with similar viscosity and surface tension but varying relaxation time, impacting on glass and hydrophobic paper substrates, with the spreading dynamics recorded by a three-camera high-speed imaging system.

We find that viscoelastic droplets follow the same leading-order inertia--capillary asymptotic scalings as Newtonian fluids. 
However, when the maximal spreading diameter of viscoelastic droplets is normalized and compared to Newtonian predictions at the same impact conditions, deviations emerge that can be attributed to fluid elasticity. 
These deviations are non-monotonic in the Deborah number and attain their maximum when the elastic relaxation time is comparable to the spreading timescale, i.e.,\ $\mathrm{De}\approx 1$.
When viscous dissipation dominates the impact, i.e., Oh$\gtrsim 0.1$, as indicated by a viscous boundary-layer thickness exceeding the spreading lamella thickness, elastic effects are masked, and the spreading reverts towards Newtonian behavior.
These results support the general trend predicted by our scaling argument: a reduced-order extension of classical energy balance that incorporates viscoelastic effects through a single correction factor, $\alpha(\mathrm{De})$. 
That is, increasing polymer concentration does not necessarily increase the deviation from Newtonian behavior; instead, measurable viscoelastic effects arise only within a narrow window where the elastic and spreading timescales are comparable. 
The framework presented here provides a direct route for incorporating viscoelasticity into well-established droplet-impact correlations and offers a basis for extending current impact models beyond purely Newtonian fluids.

Finally, future work should test the validity of the scaling argument at the opposite end of the Deborah-number range, to determine whether the predicted trend is indeed symmetric. The investigation could also be extended to resolve more precisely the transition to the viscous-dominated regime, in which elastic energy is both stored and dissipated.

The data that support the findings of this article are publicly available \cite{zenit2026}.

\begin{acknowledgments} 
We thank L. Kramer for her help in conducting initial experiments. O.A. was supported by a Fulbright Postdoctoral Fellowship. D.W. was partially supported by a UTRA Fellowship. R.Z. acknowledges the funding from the Petroleum Research Fund (Grant Number 66922-ND9) and Hazeltine Innovation Award.
\end{acknowledgments}

\appendix*
\section{Correlations for maximum droplet spread for Newtonian liquids}\label{app:correlations}
\begin{figure}
    \centering
    \includegraphics[width=\linewidth]{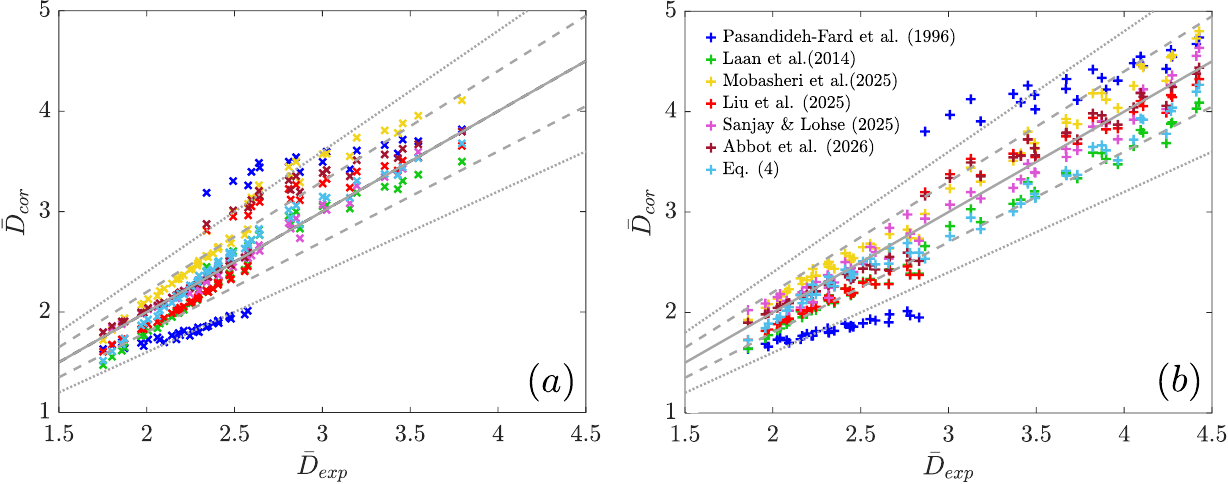}
    \caption{Comparison of predicted, $\bar{D}_{cor}$, and measured, $\bar{D}_{{exp}}$, maximum spreading diameters for the Newtonian data in this study: (a) paper substrate and (b) glass substrate.
    The solid line indicates perfect agreement, while the dashed and dotted lines indicate deviations of $\pm10\%$ and $\pm20\%$, respectively. Predictions are shown for the correlations of \citet{pasandidehfard_capillary_1996,laan_maximum_2014,mobaseri_maximum_2025,liu_maximum_2025,sanjay_unifying_2025,abbot_kinematic_2026}, together with the correlation proposed in \cref{eq:pade_new}.}
    \label{fig:N_correlations}
\end{figure}

Several correlations have been proposed to predict the maximum spreading diameter of impacting Newtonian droplets.
As discussed in \cref{sec:spread}, isolating the effect of viscoelasticity requires a reliable Newtonian reference against which the measured spreading can be compared.
To identify a suitable benchmark for the present experimental conditions, we tested several published correlations against our Newtonian data in \cref{fig:N_correlations}. 
The different correlations reproduce the overall magnitude of the maximum spreading but exhibit varying degrees of systematic over- or under-prediction, with the agreement also depending on the substrate and impact conditions.
In contrast, the correlation introduced in \cref{eq:pade_new} remains close to the measured spreading for both substrates and captures the Newtonian data without a systematic bias.
We therefore use it as the Newtonian benchmark throughout the present study, providing a consistent reference for both substrates. 
Using this baseline, we allow for deviations associated with viscoelasticity to be distinguished from deviations arising from the underlying Newtonian spreading correlation.


%

\end{document}